\documentclass[sigconf]{acmart}

\usepackage{subfigure}
\usepackage{multirow}
\usepackage{algorithm,algorithmic}
\usepackage{enumitem}
\usepackage{color,xspace}
\usepackage{xcolor}
\usepackage{soul,ulem}

\usepackage{tcolorbox}  % 文本加框
\tcbuselibrary{breakable}   % 配合文本加框，使得框可以跨页

\newcommand{\name}{SampleLLM\xspace}

\AtBeginDocument{%
  }

\setcopyright{acmlicensed}
\copyrightyear{2018}
\acmYear{2018}
\acmDOI{XXXXXXX.XXXXXXX}

\acmConference[WWW '25]{the ACM Web Conference 2025}{April 28--May 02,
  2025}{Sydney, Australia.}

\acmISBN{978-1-4503-XXXX-X/18/06}

\acmSubmissionID{xxx}

\begin{document}
\title{SampleLLM: Optimizing Tabular Data Synthesis \\ in Recommendations}

\author{Jingtong Gao}
\authornote{These authors contributed equally to this work.}
\affiliation{%
  \institution{City University of Hong Kong}
  \city{Hong Kong}
  \country{China}}
\email{jt.g@my.cityu.edu.hk}

\author{Zhaocheng Du}
\authornotemark[1]
\affiliation{%
  \institution{Huawei Noah's Ark Lab}
  \city{Shenzhen}
  \country{China}}
\email{zhaochengdu@huawei.com}

\author{Xiaopeng Li}
\affiliation{%
  \institution{City University of Hong Kong}
  \city{Hong Kong}
  \country{China}}
\email{xiaopli2-c@my.cityu.edu.hk}

\author{Yichao Wang}
\affiliation{%
  \institution{Huawei Noah's Ark Lab}
  \city{Shenzhen}
  \country{China}}
\email{wangyichao5@huawei.com}

\author{Xiangyang Li}
\affiliation{%
  \institution{Huawei Noah's Ark Lab}
  \city{Shenzhen}
  \country{China}}
\email{lixiangyang34@huawei.com}

\author{Huifeng Guo}
\affiliation{%
  \institution{Huawei Noah's Ark Lab}
  \city{Shenzhen}
  \country{China}}
\email{huifeng.guo@huawei.com}

\author{Ruiming Tang}
\authornote{Corresponding authors.}
\affiliation{%
  \institution{Huawei Noah's Ark Lab}
  \city{Shenzhen}
  \country{China}}
\email{tangruiming@huawei.com}

\author{Xiangyu Zhao}
\authornotemark[2]
\affiliation{%
  \institution{City University of Hong Kong}
  \city{Hong Kong}
  \country{China}}
\email{xianzhao@cityu.edu.hk}

\renewcommand{\shortauthors}{Gao et al.}

\begin{abstract}
Tabular data synthesis is crucial in machine learning, yet existing general methods-primarily based on statistical or deep learning models-are highly data-dependent and often fall short in recommender systems. This limitation arises from their difficulty in capturing complex distributions and understanding complicated feature relations from sparse and limited data, along with their inability to grasp semantic feature relations. Recently, Large Language Models (LLMs) have shown potential in generating synthetic data through few-shot learning and semantic understanding. However, they often suffer from inconsistent distribution and lack of diversity due to their inherent distribution disparity with the target dataset. To address these challenges and enhance tabular data synthesis for recommendation tasks, we propose a novel two-stage framework named \name to improve the quality of LLM-based tabular data synthesis for recommendations by ensuring better distribution alignment. In the first stage, \name employs LLMs with Chain-of-Thought prompts and diverse exemplars to generate data that closely aligns with the target dataset distribution, even when input samples are limited. The second stage uses an advanced feature attribution-based importance sampling method to refine feature relationships within the synthetic data, reducing any distribution biases introduced by the LLM. Experimental results on three recommendation datasets, two general datasets, and online deployment illustrate that \name significantly surpasses existing methods for recommendation tasks and holds promise for a broader range of tabular data scenarios.

\end{abstract}

%%
%% The code below is generated by the tool at http://dl.acm.org/ccs.cfm.
%% Please copy and paste the code instead of the example below.
%%
\begin{CCSXML}
<ccs2012>
   <concept>
       <concept_id>10002951.10003227.10003351</concept_id>
       <concept_desc>Information systems~Data mining</concept_desc>
       <concept_significance>500</concept_significance>
       </concept>
 </ccs2012>
\end{CCSXML}

\ccsdesc[500]{Information systems~Data mining}

%%
%% Keywords. The author(s) should pick words that accurately describe
%% the work being presented. Separate the keywords with commas.
\keywords{Recommender System, Tabular data generation, Large Language Model}

% \received{18 November 2024}
% \received[revised]{12 March 2009}
% \received[accepted]{5 June 2009}

%%
%% This command processes the author and affiliation and title
%% information and builds the first part of the formatted document.
\maketitle

\section{Introduction}

Tabular data is integral to a wide array of machine learning applications across sectors like e-commerce and healthcare, underscoring its foundational importance~\cite{habibi2023imbalanced,mcelfresh2024neural,goel2023role,bourou2021review,fayaz2022deep}. This widespread reliance amplifies the urgent demand for high-quality synthetic tabular data, particularly in recommender systems, where data sparsity poses significant challenges~\cite{hernadez2023synthetic,qian2023synthcity,bourou2021review}. Insufficient quality and volume of datasets critically impair the performance and efficiency of machine learning models~\cite{borisov2022language}.

Traditional tabular data synthesis methods~\cite{bansal2022systematic,figueira2022survey,sufi2024addressing}, focusing on a wide range of general tasks, primarily rely on statistical and deep learning models, which are heavily data-dependent and effective with abundant data availability~\cite{fonseca2023tabular}. However, when applied to recommendation tasks, these methods not only struggle in scenarios characterized by inherent data sparsity and scarcity but also fail to capture the semantic relationships between features, which is proven to be more and more crucial in recommendation modeling~\cite{liu2024multimodal,li2023text}. Consequently, there is a pressing need for innovative approaches capable of generating high-quality synthetic tabular data from minimal input while understanding semantic feature relations for recommendation tasks.

\begin{figure}[t]
    \centering
    \includegraphics[width=0.45\linewidth]{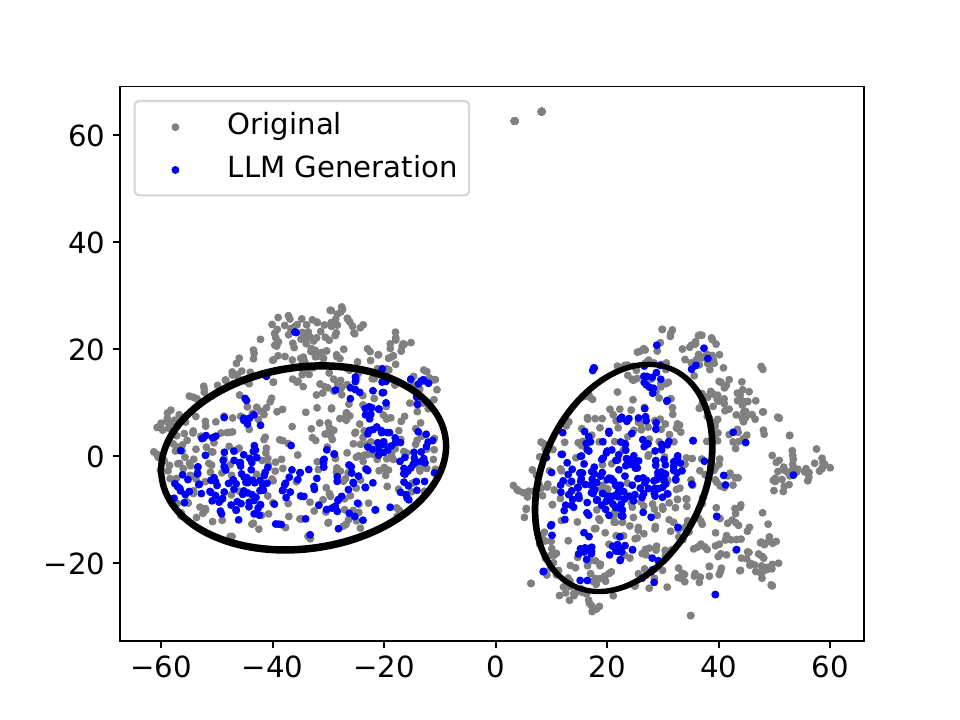}
    \vspace{-4mm}
    \caption{Visualization of LLM-generated and original tabular samples on the HELOC dataset reveals that the synthetic data lacks distribution alignment and diversity, clustering around a few centers within the original data distribution.}
    \vspace{-4mm}
    \label{fig:introduction}
\end{figure}

The advent of Large Language Models (LLMs) has marked a shift in possibilities~\cite{xu2024large}, offering new capabilities through few-shot learning and deep semantic understanding~\cite{cahyawijaya2024llms,dong2022survey,liu2024moe}. Despite their promise in synthetic data generation, LLMs frequently encounter challenges with inconsistent distributions and inaccurately modeled feature relationships~\cite{gao2023autotransfer, liu2024multifs}, both of which are critical for effective recommendation modeling~\cite{song2022autoassign,zhao2018deep, zhao2018recommendations}. These challenges arise from an intrinsic distribution mismatch between LLMs' inherent knowledge and target datasets. Moreover, when using limited exemplars, a simplistic random selection for few-shot learning might overlook important regions of the original dataset, leading to reduced output diversity, as demonstrated in Figure~\ref{fig:introduction}.

To address these challenges, we propose \name, a novel two-stage framework designed to enhance the quality of LLM-based tabular data synthesis in recommendation tasks. The first stage utilizes LLMs with Chain-of-Thought prompts~\cite{wang2023plate} and a curated selection of diverse exemplars to improve semantic understanding and feature relation modeling~\cite{lin2023autodenoise} in generating synthetic data with limited input samples. The second stage further integrates a feature attribution-based importance sampling technique to refine the synthetic data, reducing any distribution biases introduced by LLMs by employing a semi-independence assumption for feature interactions that could streamline computation while preserving the essential characteristics and relationships within the dataset.

In summary, our contribution could be summarized as follows:

\begin{itemize}[leftmargin=*]
    \item To the best of our knowledge, this is the first method to consider distribution alignment for LLM-based tabular data synthesis in recommendations.
    \item We propose a comprehensive framework \name that combines a feature attribution-based importance sampling strategy for distribution alignment with advanced few-shot LLM generation techniques. This integration significantly enhances the utility and distribution consistency of the generated tabular data.
    \item Experiments conducted on three widely used recommendation datasets, two general datasets, and through online deployment reveal that the synthetic tabular data generated by \name not only surpasses existing baselines in recommendation tasks but also shows promise for broader application in general tasks.
\end{itemize}
\section{Framework}
In this section, we first describe the problem formulation of the tabular data synthesis, and then provide an overview of \name and detail its key components. 

\subsection{Problem Formulation}\label{sec:problem}
The task of tabular data synthesis involves generating synthetic tabular data that closely resemble those in a given dataset. The synthetic data is used for various purposes, including augmenting training datasets and conducting experiments without exposing sensitive information. Formally, let $\mathcal{D}_o = \{\boldsymbol{x}_1, \boldsymbol{x}_2, \ldots, \boldsymbol{x}_N\}$ represent the original dataset with $N$ samples, where each $\boldsymbol{x}_i$ is a sample with $K$ attributes and one label $y_i$, i.e., $\boldsymbol{x}_i = \{x_i^1, x_i^2, \ldots, x_i^K, y_i\}$. The objective is to develop a generative model $G$ that produces a synthetic dataset with $T$ samples $\mathcal{D}_s = \{\boldsymbol{x}_1', \boldsymbol{x}_2', \ldots, \boldsymbol{x}_T'\}$ such that $\mathcal{D}_s$ exhibits similar statistical properties and distributional characteristics as $\mathcal{D}_o$ and could be applied to various tasks with similar performance. These requirements can be formalized as follows:

\begin{itemize}[leftmargin=*]
    \item \textbf{Distributional Similarity}: The synthetic dataset $\mathcal{D}_s$ should follow the similar distribution as the original dataset $\mathcal{D}_o$. This can be expressed as:
    \begin{equation}
        P_{\mathcal{D}_s}(X) \approx P_{\mathcal{D}_o}(X),
    \end{equation}
    where $P_{\mathcal{D}_s}(X)$ and $P_{\mathcal{D}_o}(X)$ denote the probability distributions of the synthetic and original datasets, respectively.
    \item \textbf{Utility}: The synthetic dataset $\mathcal{D}_s$ should be useful for the same downstream tasks as the original dataset $\mathcal{D}_o$. If $\mathcal{D}_s$ is used to train a machine learning model $\mathcal{M}$, the model's performance on downstream tasks should closely match that of a model trained on the original dataset $\mathcal{D}_o$. Formally, this can be expressed as:
    \begin{equation}\label{equ:alter}
    U(\mathcal{D}_s, \mathcal{M}) \approx U(\mathcal{D}_o, \mathcal{M}),
    \end{equation}
    where $U(\mathcal{D}_o, \mathcal{M})$ denotes the utility of $\mathcal{D}_o$ for model $\mathcal{M}$. For simplicity, this aspect is referred to as \textbf{Machine Learning Efficacy (MLE) utility}~\cite{ctgan,solatorio2023realtabformer} in this paper. In addition, in terms of data enhancement~\cite{figueira2022survey,chatterjee2022enhancement}, we aim for the machine learning model $\mathcal{M}$ trained on the combined dataset $\mathcal{D}_o + \mathcal{D}_s$ to achieve performance on downstream tasks that is comparable to or surpasses that of a model trained exclusively on the original dataset $\mathcal{D}_o$.
    Formally, this can be expressed as:
    \begin{equation}\label{equ:cross}
    U(\mathcal{D}_o + \mathcal{D}_s, \mathcal{M}) \gtrsim U(\mathcal{D}_o, \mathcal{M})
    \end{equation}

    For simplicity, this aspect is referred to as \textbf{augumentation utility} in this paper.
\end{itemize}

By ensuring these properties, the synthetic dataset $\mathcal{D}_s$ can effectively supplement or substitute for the original dataset $\mathcal{D}_o$ in various applications, providing a valuable tool for data argumentation and analysis while preserving privacy and confidentiality.

\subsection{\name Overview}
\begin{figure*}[t]
    \centering
    \includegraphics[width=0.95\linewidth]{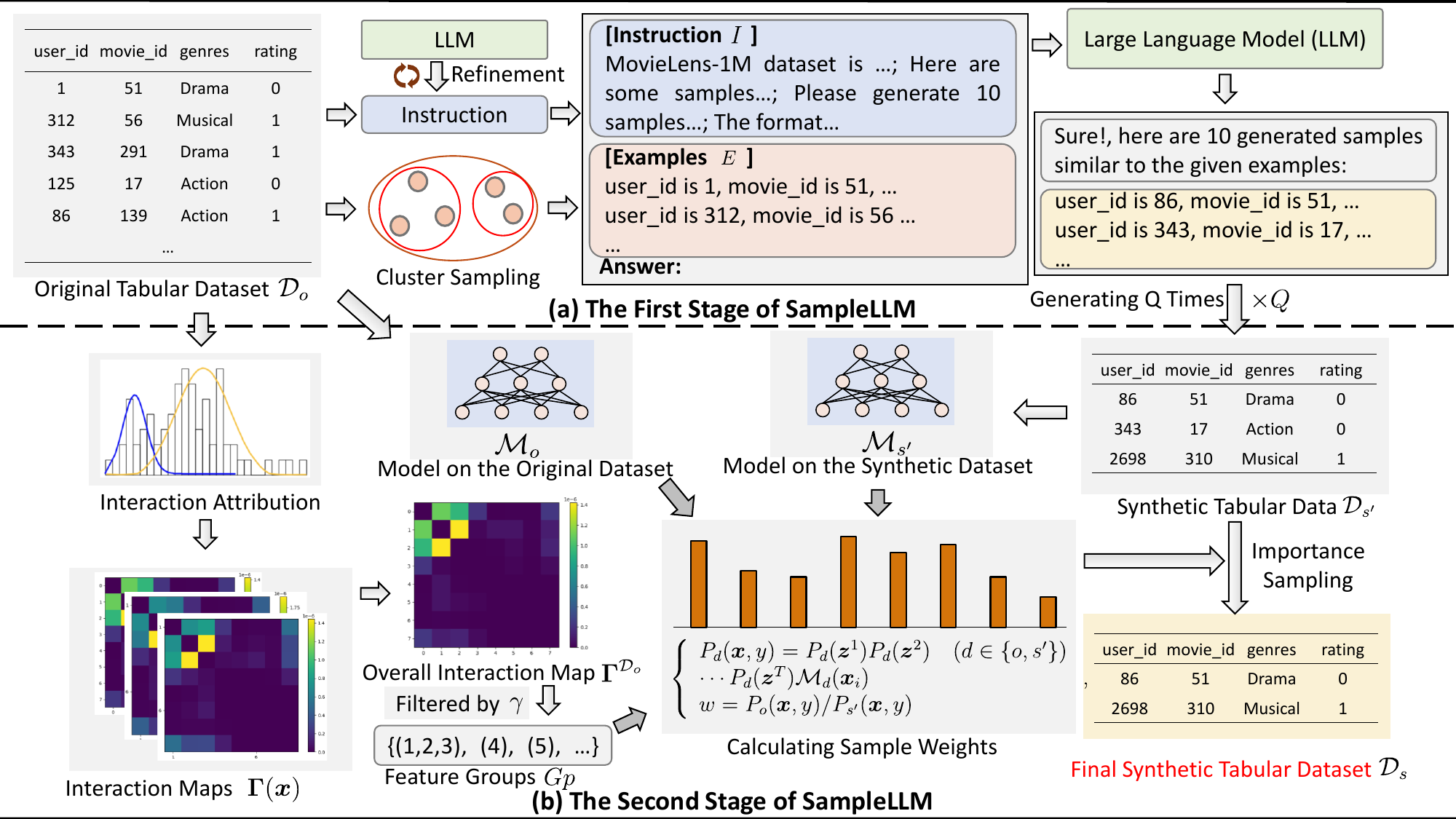}
    \vspace{-2mm}
    \caption{The overall structure of \name. (a) In the first stage, a manually designed instruction and $a$ samples extracted with clustering sampling are used as inputs to the LLM, which generates $b$ synthetic samples. This process is repeated $Q$ times. (b) In the second stage, a novel feature attribution-based importance sampling method is employed on the synthetic samples. }
    \label{fig:overview}
    \vspace{-2mm}
\end{figure*}

As illustrated in Figure~\ref{fig:overview}, \name consists of two stages. In the first stage, the designed instruction and sampled exemplars from the original dataset $\mathcal{D}_o$ are selected to serve as the input for the LLM, producing synthetic tabular data $\mathcal{D}_{s'}$. In the second stage, a novel feature attribution-based importance sampling operation is performed to achieve further feature relation modeling and distributional alignment between the synthetic and original datasets, resulting in the final synthetic dataset $\mathcal{D}_{s}$.

Specifically, in the first stage, an instruction $I$ is designed to explain the data generation task to the LLM. This instruction is first selected from a set of manually crafted instructions based on performance criteria, and then refined by LLM with official documents and a small number of samples in a Chain-of-Thought (CoT) manner to extract key information. Following this, a cluster sampling method is applied to the original tabular data to generate $a$ exemplars represented by $E$. This ensures that the data generated by the LLM is diverse and closely resembles the original distribution. By combining the instruction and exemplars as inputs, the LLM generates $b$ synthetic samples using a few-shot learning strategy. This process is repeated $Q$ times to produce a total of $bQ$ synthetic samples, where $a$, $b$, and $Q$ are hyper-parameters. In the second stage, an innovative importance sampling method is applied to the synthetic dataset $\mathcal{D}_{s'}$ to further align its distribution with that of the original data $\mathcal{D}_o$. The weight of each sample is determined by the joint distribution probability of the features and labels under a semi-independent hypothesis where the non-independent feature groups are obtained through the overall interaction map of the original dataset and the discriminant probabilities of different labels are given by a predictive model trained on the corresponding dataset.

\begin{figure}[t]
    \centering
    \includegraphics[width=0.85\linewidth]{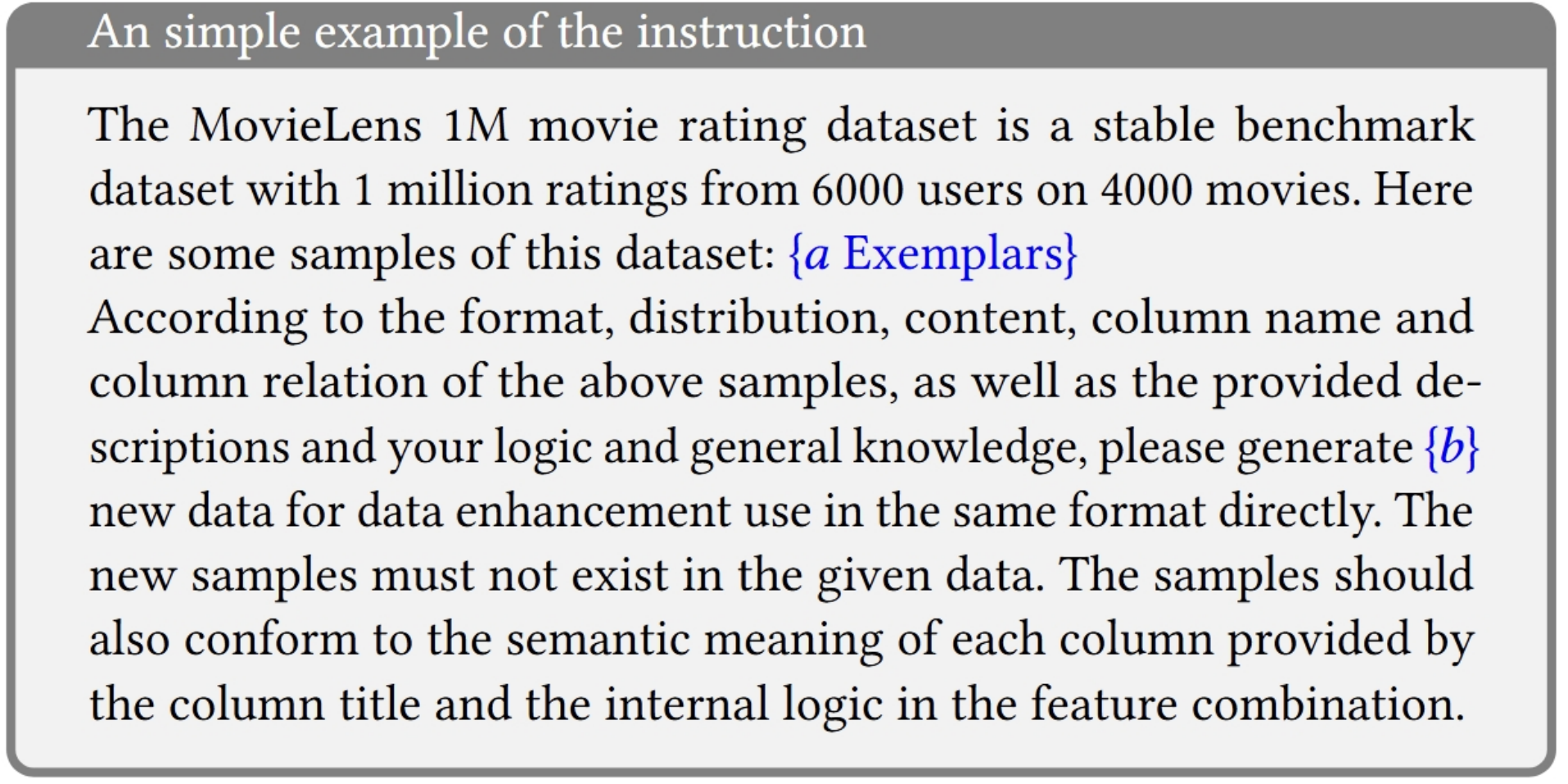}
    \vspace{-4mm}
    \caption{A simplified instruction example.}
    \vspace{-5mm}
    \label{fig:exemplars}
\end{figure}

\subsection{Designed Instruction}\label{sec:instruction}

Traditional statistical and deep learning models are predominantly data-driven, necessitating substantial volumes of samples to accurately model data distributions. This reliance significantly limits their effectiveness when dealing with recommendation tasks with small and sparse datasets, where synthetic data generation becomes crucial. Furthermore, when it comes to textual features, these models typically use ID encodings as input, hindering their ability to capture the semantic associations between features. To address these limitations, we leverage the general knowledge and semantic understanding capabilities of LLMs for data synthesis via few-shot learning. A pivotal component of this approach is the precise framing of the tabular data synthesis task within the prompt, referred to as the instruction, as illustrated in Figure~\ref{fig:exemplars}. Here, we identify the most effective instruction from a set of manually crafted options tailored to each dataset. In our study, the final instruction for each dataset is selected from five custom-designed instructions. Specifically, for each dataset, 1\% synthetic samples are generated and incorporated into the training set. The instruction that delivers the best performance is then established as the standard for that dataset. To further enhance the instruction with relevant knowledge, we input the manually curated instruction, along with official documents $O$ and a few data samples $S$, into the LLM, which iteratively refines the instruction expression using a Chain-of-Thought process. This procedure is described as follows:
\begin{equation}\label{equ:instruction}
I_{t+1} = LLM(I_t, O, S)
\end{equation}
where the maximum value of iteration times $t$ is fixed at 5. For simplicity, we denote the final output instruction as $I$.

% \begin{tcolorbox}[boxsep=0mm,
%             colframe=gray,
%             width=1\linewidth,
%             arc=1mm, 
%             auto outer arc,
%             title={\small An simple example of the instruction},]
%     \small The MovieLens 1M movie rating dataset is a stable benchmark dataset with 1 million ratings from 6000 users on 4000 movies. Here are some samples of this dataset: \nodecolor{$a$ Exemplars}\\
%     \small According to the format, distribution, content, column name and column relation of the above samples, as well as the provided descriptions and your logic and general knowledge, please generate \nodecolor{$b$} new data for data enhancement use in the same format directly. The new samples must not exist in the given data. The samples should also conform to the semantic meaning of each column provided by the column title and the internal logic in the feature combination.
%     \end{tcolorbox}

\subsection{Selecting Exemplars}

As the key to few-shot learning, exemplars are crucial for providing LLMs with a small number of examples that standardize the format, content, and distribution of LLM-generated samples. However, current studies~\cite{borisov2022deep} often select exemplars randomly from the original dataset. This approach may fail to capture the diverse distribution of the original data, especially since the number of exemplars is limited by the maximum input tokens of the LLM. Furthermore, the intrinsic distribution differences between LLM's inherent knowledge and target dataset exacerbate this issue. Consequently, as shown in Figure~\ref{fig:introduction}, significant distribution discrepancies are observed between the generated and original tabular samples.

To address these problems, two alignment methods are applied in \name. The first method aims to maintain sample diversity during exemplar selection to avoid distribution concentration, as detailed below. The second method discussed further in Section~\ref{sec:importance}, leverages a novel feature attribution-based importance sampling method to model critical feature relations and address the distribution differences caused by the LLM.

To ensure exemplars reflect the diverse characteristics of the original data and mitigate the sample aggregation issue in LLM output, a simple clustering-based sampling method is first proposed in \name for exemplar selection. This approach is mathematically represented as follows:
\begin{equation}\label{equ:exemplar}
\left\{
\begin{aligned}
    &Clu = h(\mathcal{D}_o, a) = \{clu_1, clu_2, \ldots, clu_a\}\\
    &E = \{\boldsymbol{x}_i | \boldsymbol{x}_i = g(clu_i); i \in \{1,2, \ldots, a\}\}
\end{aligned}
\right.
\end{equation}

Here, $a$ clusters $\{clu_1, clu_2, \ldots, clu_a\}$ are derived from the original dataset $\mathcal{D}_o$ using a clustering algorithm $h$—specifically, the K-means method. From each cluster, one exemplar is selected using method $g$—in this case, random selection—to form an exemplar set $E$. This approach ensures that each group of exemplars contains samples from diverse regions of the original dataset's distribution, thereby enhancing diversity and achieving better alignment with the distribution characteristics of the original data.

As illustrated in Figure~\ref{fig:overview} (a), due to the maximum output token limitations in LLMs, the entire first stage of \name will iterate $Q$ times, with $b$ samples generated by LLM each time, where $Q$ and $b$ are hyper-parameters. After generation, the generated $bQ$ samples will be formatted in tabular form and aggregated together to form a temporary synthetic tabular dataset $\mathcal{D}_{s'}$.

\subsection{Feature Attribution-based Importance Sampling}\label{sec:importance}

Although cluster sampling methods are designed to mitigate discrepancies in data distributions, inherent differences caused by the LLM's input-output processing still result in a distribution gap and differentiated feature interaction frequency between the generated and the original data. This gap poses significant challenges in recommendation scenarios, where feature associations serve as a foundational modeling basis. To address these issues, \name employs a feature attribution-based importance sampling method to further refine the feature relation and alignment of distribution in $\mathcal{D}_{s'}$ with that in $\mathcal{D}_o$~\cite{benchmark, janizek2020explaining}. Our method is proposed because directly calculating the exact importance weights for tabular samples is challenging due to the high dimensionality of tabular data, which complicates the estimation of the joint distribution of the features and labels in samples:
\begin{equation}
    P(\boldsymbol{x}_i) = P(x_i^1, x_i^2, \ldots, x_i^K) P(y_i \mid x_i^1, x_i^2, \ldots, x_i^K)
\end{equation}

One straightforward approach to simplify the calculation is to assume independence among all features:
\begin{equation}
    P(\boldsymbol{x}_i) = P(x_i^1)P(x_i^2)\ldots P(x_i^K) P(y_i \mid x_i^1, x_i^2, \ldots, x_i^K)
\end{equation}

However, this assumption often diverges significantly from reality, as it is generally impractical to assume complete independence among all features in most datasets. Consequently, this approach may lead to inaccuracies in estimating distribution probabilities, thereby complicating the data alignment process.

To mitigate these issues, \name adopts a semi-independence assumption where feature fields containing significant interactions (e.g., fields 1 and 8) are considered non-independent, while others are treated as independent. This balanced approach reduces computational overhead while minimizing calculation errors, enabling the computation of importance weights based on frequency statistics:
\begin{equation}\label{equ:interaction}
    P(\boldsymbol{x}_i) = P(x_i^1, x_i^8)P(x_i^2)\ldots P(x_i^K) P(y_i \mid x_i^1, x_i^2, \ldots, x_i^K)
\end{equation}

\subsubsection{Feature Interaction Extraction}

To determine the distribution probability of a sample within a specific dataset, as specified in Equation~\eqref{equ:interaction}, it is essential to identify significant feature interactions. This paper introduces a feature attribution-based method for extracting such interactions.

For a dataset $\mathcal{D}$ linked to a particular prediction task and a deep learning-based predictive model $f$, we could quantify the importance of each feature to the output by approximating model performance degradation after ablating this feature with 1st-order Taylor approximation~\cite{sundararajan2016gradients} as follows:

\begin{equation}
\text {EG}_i(\boldsymbol{x}):=\underset{\boldsymbol{x}^{\prime} \sim \mathcal{D}^b}{\mathbb{E}}\left[\left(x_i-x_i^{b}\right) \frac{\delta f\left(\boldsymbol{x}\right)}{\delta x_i}\right]
\end{equation}

where $x_i$ are features in $\boldsymbol{x}$, $x_i^{b}$ are non-informative features like zero feature or random feature and $\mathcal{D}^b$ is typically a randomly selected subset of $\mathcal{D}$ or a single sample containing all-zero values~\cite{erion2019learning,janizek2021explaining}. This approach quantifies the importance of each feature to the output by approximating model performance change after setting an informative feature to non-informative.

To attribute feature interactions, we extend 1st-order Taylor expansion to 2nd-order, capturing the importance of feature $i$ to the importance of feature $j$. The resulting expression is:

\begin{equation}\label{equ:gamma_1}
\begin{aligned}
\boldsymbol{\Gamma}_{i, j}(x) & = \underset{\boldsymbol{x} \sim \mathcal{D}^b}{\mathbb{E}}\left[\left(x_i-x_i^{b}\right)\left(x_j-x_j^{b}\right) \frac{\partial^2 f\left(\boldsymbol{x}^{b}\right)}{\partial x_i \partial x_j}\right]
\end{aligned}
\end{equation}

Intuitively, this method measures feature interaction strength as approximated feature importance change when switching interaction from informative to non-informative.

By applying this interaction attribution method to sample samples from the training set, we could capture the complex interactions between feature fields across the entire dataset. Therefore, summing the absolute values of interaction maps $\boldsymbol{\Gamma}(\boldsymbol{x})$ for all samples yields a comprehensive matrix that highlights the most significant feature interactions in the whole dataset, irrespective of whether their impacts on the output are positive or negative:
\begin{equation}\label{equ:gamma_2}
\boldsymbol{\Gamma}^{\mathcal{D}_o} = \sum_{\boldsymbol{x} \sim \mathcal{D}_o} \lvert\boldsymbol{\Gamma}(\boldsymbol{x})\rvert
\end{equation}

Interaction values above a threshold, defined by a hyper-parameter $\gamma$, are then clustered to delineate independent and non-independent feature groups, further refining data alignment:
\begin{equation}\label{equ:group}
\left\{
\begin{aligned}
&F_{\gamma} = \gamma \cdot max(\boldsymbol{\Gamma}^{\mathcal{D}_o})\\
&Pairs = \text{argwhere}_{\boldsymbol{\Gamma}^{\mathcal{D}_o}}(\boldsymbol{\Gamma}^{\mathcal{D}_o}>F_{\gamma})\\
&Gp = \text{Merge}(Pairs)
\end{aligned}
\right.,
\end{equation}
where the $Merge$ function aggregates all pairs with overlapping feature fields to form a feature group $\boldsymbol{z}^t$ (e.g., $\boldsymbol{z}^1 = \{x^1, x^3, x^8\}$) with several feature fields, resulting in $T$ groups $Gp=\{\boldsymbol{z}^1, \ldots, \boldsymbol{z}^t, \ldots, \boldsymbol{z}^T\}$ where $T$ is determined by the aggregation process, ensuring $T \leq K$.

In this study, our objective is to align the distribution of the dataset $\mathcal{D}_{s'}$ with that of the original dataset $\mathcal{D}_o$. Consequently, the feature groups are derived from $\mathcal{D}_o$ and are utilized for calculating the distribution probability of samples in both $\mathcal{D}_o$ and $\mathcal{D}_{s'}$. This methodology facilitates a refined understanding of the underlying feature interactions, thus enhancing model interpretability and improving alignment between datasets.

\subsubsection{Obtaining Discriminant Probability}

As a dataset, each tabular sample contains not only a set of features but also a label. Therefore, another critical challenge in calculating the probability of a sample is obtaining the discriminant probability $P(y_i \mid x_i^1, x_i^2, \ldots, x_i^K)$ for each sample $\boldsymbol{x}_i=\{x_i^1, x_i^2, \ldots, x_i^K, y_i\}$. In \name, this is achieved by training a task-specific predictive model $\mathcal{M}$ on the corresponding dataset:

\begin{equation}\label{equ:probability}
\mathcal{M}(\boldsymbol{x}_i) = P(y_i \mid x_i^1, x_i^2, \ldots, x_i^K)
\end{equation}

For a specific task (e.g., binary classification), two predictive models $\mathcal{M}_o$ and $\mathcal{M}_{s'}$ are trained to predict the sample discriminant probabilities on $\mathcal{D}_o$ and $\mathcal{D}_{s'}$, respectively.

\subsubsection{Calculating Sample Weights}
After obtaining the feature groups $Gp$ and discriminant models $\mathcal{M}_o$ and $\mathcal{M}_{s'}$, the importance weights of samples can be derived using the following formula:

\begin{equation}\label{equ:weights}
\left\{\begin{array}{l}
P_d(\boldsymbol{x}, y)=P_d(\boldsymbol{z}^1) P_d(\boldsymbol{z}^2) \quad(d \in\{o, s'\}) \\
\cdots P_d(\boldsymbol{z}^T) 
 \mathcal{M}_d(\boldsymbol{x}_i) \\
w=P_o(\boldsymbol{x}, y) / P_{s'}(\boldsymbol{x}, y)
\end{array}\right.,
\end{equation}
where the first term is an alternative expression of Equation~\eqref{equ:interaction}, and the second term is used to calculate the importance weights based on the sample's distribution probability in both the original and the synthetic datasets. $P_d()$ is the frequency of related features in dataset $d$.

Finally, after obtaining all synthetic samples' weights, the final synthetic tabular dataset $\mathcal{D}_{s}$ is created through an importance sampling on $\mathcal{D}_{s'}$. The overall pseudo-code is also provided in Appendix~\ref{sec:app-pseudo} to further clarify the overall flow of \name.

\section{Experiments}

In this section, we conduct experiments on three recommendation datasets to address the following research questions:

\begin{itemize}[leftmargin=*]
    \item \textbf{RQ1:} How does \name perform in comparison to other tabular data synthesis methods?
    \item \textbf{RQ2:} How effective is the generated tabular data when used as a replacement for the original tabular data?
    \item \textbf{RQ3:} What impact do cluster sampling and feature attribution-based importance sampling in \name have in the process of generating synthetic tabular data?
\end{itemize}

Moreover, two extensive general datasets are also applied to validate \name's effectiveness in general tabular data-related tasks. In the following sections, we first describe the experimental setup used in this study in the following subsections. Subsequently, detailed analyses are provided for each research question based on our experimental results.

% Table generated by Excel2LaTeX from sheet 'statistics'
\begin{table}[t]
  \centering
  \caption{Data Statistics.}
  \vspace{-4mm}
    \begin{tabular}{cccc}
    \toprule
    Dataset & Train & Validation  & Test \\
    \midrule
    ML-1M     &800,167    & 100,020  & 100,022 \\
    Amazon     & 658,827     & 82,353 & 82,354 \\
    Douban     & 1,348,399     & 168,549 & 168,551 \\
    HELOC      &8,367 &  1,045 & 1,047 \\
    Covertype & 464,809 & 58,101 & 58,102 \\
    \bottomrule
    \end{tabular}%
    \vspace{-6mm}
  \label{tab:statistics}%
\end{table}%

\subsection{Experimental Setup}
\subsubsection{Dataset}
We conducted experiments on three commonly used recommendation datasets with Click-Through Rate (CTR) prediction tasks~\cite{fu2023unified,liang2023mmmlp} to verify the validity of \name, i.e., ML-1M~\footnote{https://grouplens.org/datasets/movielens/1m/}, Amazon~\footnote{https://jmcauley.ucsd.edu/data/amazon/} and Douban~\footnote{https://www.kaggle.com/datasets/fengzhujoey/douban-datasetratingreviewside-information}. In addition, a binary classification dataset HELOC~\footnote{https://huggingface.co/datasets/mstz/heloc} and a multi-classification dataset Covertype~\footnote{https://archive.ics.uci.edu/dataset/31/covertype} are further applied for extended experiments to verify the potential of \name in general tasks. In accordance with previous studies~\cite{yuan2022tenrec,lian2018xdeepfm}, each dataset is divided into training/validation/test sets in an 8:1:1 ratio. Data statistics could be found in Table~\ref{tab:statistics}.

% Table generated by Excel2LaTeX from sheet 'overall-1'
\begin{table*}[t]
  \centering
  \caption{Augmentation utility. The boldface denotes the best score. The underline indicates the second-best score. The ``Original'' method indicates training without synthetic data and is not considered in scoring. $\uparrow$: higher is better; $\downarrow$: lower is better. ``\textbf{{\Large *}}'' indicates the statistically significant improvements (i.e., two-sided t-test with $p<0.05$) over the best baseline.}
  \vspace{-4mm}
  \scalebox{0.77}{
    \begin{tabular}{cccccccccccc}
    \toprule
    \multirow{2}[4]{*}{Approach} & \multicolumn{2}{c}{ML-1M} & \multicolumn{2}{c}{Amazon} & \multicolumn{2}{c}{Douban} & \multicolumn{2}{c}{HELOC} & \multicolumn{3}{c}{CoverType} \\
\cmidrule{2-12}          & AUC $\uparrow$  & Logloss $\downarrow$ & AUC $\uparrow$  & Logloss $\downarrow$ & AUC $\uparrow$  & Logloss $\downarrow$ & AUC $\uparrow$  & Logloss $\downarrow$ & Precision $\uparrow$ & Recall $\uparrow$ & F1 $\uparrow$ \\
    \midrule
    Original & 0.8173  & 0.5163  & 0.7075  & 0.4602  & 0.8016  & 0.5158  & 0.7638  & 0.6445  & 0.7395  & 0.7339  & 0.7250  \\
    PATE-GAN & 0.8154  & \underline{0.5153}  & 0.7022  & 0.4620  & 0.8010  & 0.5162 & 0.7648  & 0.6461  & 0.7388  & 0.7350  & 0.7258  \\
    ADS-GAN & 0.8147  & 0.5157  & 0.7026  & 0.4619  & 0.8016  & 0.5151 & 0.7672  & 0.6451  & 0.7414  & 0.7413  & 0.7312  \\
    CTGAN & 0.8148  & 0.5156  & 0.7027  & 0.4631  & 0.8016  &  0.5150  & 0.7677  & 0.6419  & 0.7417  & 0.7366  & 0.7295  \\
    TVAE  & 0.8143  & 0.5162  & 0.7031  & 0.4616  & 0.8019  & 0.5146   & 0.7701  & 0.6455  & 0.7422  & 0.7378  & 0.7305  \\
    TabDDPM & 0.8141  & 0.5159  & 0.7036  & 0.4619  & 0.8020 & 0.5150  & 0.7704  & 0.6424  & 0.7418  & 0.7388  & 0.7299  \\
    GReaT & 0.8153  & 0.5198  & 0.7040  & 0.4616  & 0.8021  & 0.5147  & 0.7703  & 0.6444  & 0.7423  & \textbf{0.7440 } & \underline{0.7317}  \\
    REaLTabFormer & \underline{0.8156}  & 0.5182  & \underline{0.7041}  & \underline{0.4611}  &  \underline{0.8022}   &    \underline{0.5145} & \underline{0.7707}  & \underline{0.6417}  & \underline{0.7428}  & 0.7351  & 0.7264  \\
    SampleLLM & \textbf{0.8180*} & \textbf{0.5140*} & \textbf{0.7082*} & \textbf{0.4601*} & \textbf{0.8027*}   & \textbf{0.5139*}  & \textbf{0.7732*} & \textbf{0.6403*} & \textbf{0.7445*} & \underline{0.7437}  & \textbf{0.7364*} \\
    \bottomrule
    \end{tabular}%
    }
    \vspace{-4mm}
 \label{tab:augment}%
\end{table*}%

\subsubsection{Baselines}
To verify the effectiveness of \name, a comprehensive comparison is conducted between \name and various baseline approaches:
\begin{itemize}[leftmargin=*]
    \item \textbf{CTGAN}~\cite{ctgan} utilizes a conditional generator and a mode-specific normalization method in synthetic tabular data generation to address the complexities of tabular data, including mixed data types and imbalanced categories.
    \item \textbf{TVAE}~\cite{ctgan}, a Variational AutoEncoder (VAE) adapted for synthetic tabular data generation, leverages specialized preprocessing and a modified loss function to handle mixed data types.
    \item \textbf{TabDDPM}~\cite{kotelnikov2023tabddpm} leverages a denoising diffusion probabilistic model to handle the inherent heterogeneity of tabular data, providing superior generative performance across benchmarks.
    \item \textbf{PATE-GAN}~\cite{jordon2018pate} leverages the Private Aggregation of Teacher Ensembles (PATE) framework to introduce differential privacy guarantees into the Generative Adversarial Network (GAN) setting. By modifying the GAN discriminator with the PATE mechanism, PATE-GAN ensures that the synthetic data generated maintains privacy while preserving data utility. 
    \item \textbf{ADS-GAN}~\cite{yoon2020anonymization} is designed to generate synthetic data that closely approximate the joint distribution of variables in original datasets. Utilizing a conditional Generative Adversarial Network (GAN) framework, ADS-GAN ensures data anonymization by minimizing identifiability based on the probability of re-identification. 
    \item \textbf{GReaT}~\cite{borisov2022language} leverages an auto-regressive generative LLM to synthesize realistic tabular data. This method efficiently models tabular data distributions by conditioning on any subset of features, allowing for flexible and authentic data generation.
    \item \textbf{REaLTabFormer}~\cite{solatorio2023realtabformer} generates high-quality synthetic data for both non-relational and relational datasets. It employs an autoregressive model for parent tables and a sequence-to-sequence model for related child tables with GPT-2 LLM as backbone modules, ensuring realistic data relationships and preventing data copying through target masking and statistical bootstrapping.

\end{itemize}

% Table generated by Excel2LaTeX from sheet 'overall-2'
\begin{table*}[t]
  \centering
  \caption{MLE utility. The boldface denotes the best score, and the underline indicates the second-best score. The ``Original'' method indicates training with the original training set and is not considered in scoring. $\uparrow$: higher is better; $\downarrow$: lower is better. ``\textbf{{\Large *}}'' indicates the statistically significant improvements (i.e., two-sided t-test with $p<0.05$) over the best baseline.}
  \vspace{-3mm}
  \scalebox{0.77}{
    \begin{tabular}{cccccccccccc}
    \toprule
    \multirow{2}[4]{*}{Approach} & \multicolumn{2}{c}{ML-1M} & \multicolumn{2}{c}{Amazon} & \multicolumn{2}{c}{Douban} & \multicolumn{2}{c}{HELOC} & \multicolumn{3}{c}{Covertype} \\
\cmidrule{2-12}         & AUC $\uparrow$  & Logloss $\downarrow$ & AUC $\uparrow$  & Logloss $\downarrow$ & AUC $\uparrow$  & Logloss $\downarrow$ & AUC $\uparrow$  & Logloss $\downarrow$ & Precision $\uparrow$ & Recall $\uparrow$ & F1 $\uparrow$ \\
    \midrule
    Original & 0.8173  & 0.5163  & 0.7075  & 0.4602  &   0.8016 
   &  0.5158  & 0.7638  & 0.6445  & 0.7395  & 0.7339  & 0.7250  \\
    PATE-GAN & 0.4944  & 0.9450  & 0.5800  & 0.5001  & 0.7560 
 & \underline{0.5600}   & 0.6215  & 0.6906  & 0.7126  & \underline{0.7152}  & 0.7006  \\
    ADS-GAN & 0.5590  & \underline{0.6771}  & 0.5802  & 0.5819  &  0.7548  & 0.5624  & 0.7211  & 0.9640  & 0.7069  & 0.7115  & 0.6953  \\
    CTGAN & 0.5602  & 0.6774  & 0.5775  & 0.6092  & 0.7551  & 0.5603 
  & 0.6289  & 0.6817  & 0.7055  & 0.7132  & 0.6991  \\
    TVAE  & \underline{0.5622}  & 0.6776  & 0.5828  & 0.5981  &0.7552    &  0.5610  & 0.7627  & 0.5856  & 0.7070  & 0.7091  & 0.6906  \\
    TabDDPM & 0.5334  & 0.6806  & 0.5853  & 0.5005  & 0.7556   & 0.5603 & 0.7388  & 0.6028  & 0.7105  & 0.7132  & 0.6993  \\
    GReaT & 0.5596  & 0.8181  & 0.5881  & \underline{0.4986}  & 0.7562  & 0.5615 & 0.7444  & 0.5987  & 0.7111  & 0.7134  & 0.7010  \\
    REaLTabFormer & 0.5574  & 0.7718  & \underline{0.5900}  & 0.4992  & \underline{0.7567}  &  0.5604  & \underline{0.7628}  & \underline{0.5826}  & \underline{0.7175}  & 0.7103  & \underline{0.7023}  \\
    SampleLLM & \textbf{0.5668*} & \textbf{0.6758*} & \textbf{0.5922*} & \textbf{0.4977*} & \textbf{0.7590*}  & \textbf{0.5571*}  & \textbf{0.7652*} & \textbf{0.5799*} & \textbf{0.7196*} & \textbf{0.7159*} & \textbf{0.7086*} \\
    \bottomrule
    \end{tabular}%
    }
    \vspace{-3mm}
  \label{tab:MLE}%
\end{table*}%

\subsubsection{Implementation Details}
For ML-1M, Amazon, Douban and HELOC datasets, the widely used metrics of AUC and Logloss are deployed for evaluation~\cite{kirasich2018random,guo2017deepfm}. For Covertype whose task is multi-class classification, the widely used metrics of weighted Precision, Recall, and F1 score~\cite{grandini2020metrics} are deployed for evaluation. For a fair comparison and clearly illustrate the effectiveness of each module in \name. A simple Deep Neural Network (DNN) is selected as the predictive model for all tasks. Specifically, for different datasets and tasks, we first carried out a grid search for network parameters to ensure prediction performance, in which the search range for the number of layers is \{1-4\} and the search range for the number of neurons per layer is \{16,32,64,128\}. The simple grid search is also applied in selecting other hyper-parameters such as learning rate and dropout rate in training. Meanwhile, the INT4 quantized version of Llama-3-70b-instruct-awq~\cite{llama3modelcard} is applied as the LLM backbone for the first stage of \name. In addition, we ran each experiment 10 times and reported the average performance.

\subsection{Overall Performance (RQ1, 2)}
This section gives overall performance comparisons between \name and various baselines on the augmentation utility and MLE utility as illustrated in Section~\ref{sec:problem} to answer the RQ1 and RQ2.

\subsubsection{Augmentation Utility}

As illustrated in Equation~\eqref{equ:cross}, the utility of synthetic data could be evaluated by the data enhancement effect after injecting synthetic data into the original data~\cite{figueira2022survey,chatterjee2022enhancement}. Specifically, we generate 10\% synthetic data for the training set using different baselines, and integrate them with the original training set for training. The performance of the final predictive model could, therefore, serve as evidence to measure the utility of the synthetic data. The experimental results are presented in Table~\ref{tab:augment}. A case study is also provided in Appendix~\ref{sec:case} for visualization analysis.

From Table~\ref{tab:augment} we can conclude that:
\begin{itemize}[leftmargin=*]
    \item For baselines without LLM in modeling, TabDDPM and TVAE outperform GAN-based models in most cases, verifying the effectiveness of more detailed modeling for distribution alignment in diffusion probabilistic models and VAE models. 
    \item For the ML-1M, Amazon, and Douban datasets, traditional models generally struggle to match the performance seen on the original data. However, they tend to perform better on the HELOC and Covertype datasets. This disparity highlights the importance of complex feature associations in recommendation modeling, which traditional models fail to capture due to their inability to understand semantic relationships. GReaT and REaLTabFormer outperform other baselines, showcasing LLMs' strength in modeling feature relations through semantic comprehension. Additionally, by utilizing alignment methods, \name achieves superior AUC and Logloss across all datasets.
    \item \name outperforms all baselines in most cases, illustrating the effectiveness of LLM's semantic understanding in modeling feature relations and data distributions and the effectiveness of the feature attribution-based importance sampling method in further aligning the feature correlations and data distribution of the generated data with the original data.
\end{itemize}

\subsubsection{MLE Utility}
As illustrated in Equation~\eqref{equ:alter}, the utility of synthetic data could also be evaluated by supplanting the real data in training a predictive model. The corresponding evaluation method is referred to as Machine Learning Efficacy (MLE) utility~\cite{ctgan,solatorio2023realtabformer}. Specifically, to reduce the cost of training time, we generate 10\% synthetic data for the training set using different baselines and directly use them as the training sets for model training. The performance of the final predictive model could, therefore, serve as evidence to measure the utility of the synthetic data. The experimental results are presented in Table~\ref{tab:MLE}.

From Table~\ref{tab:MLE} we can conclude that:
\begin{itemize}[leftmargin=*]
    \item Almost all methods exhibit poorer performance compared to directly using the original training set (``Original''). This result is intuitive, as synthetic data cannot fully replicate or substitute the nuances of the original data. Interestingly, \name surpasses ``Original'' on the HELOC dataset, likely due to the dataset's data scarcity. The model's convergence rate may remain similar on HELOC whether trained on the original training set or synthetic data.

    \item The performance degradation across all methods is most pronounced on the ML-1M dataset. This is likely due to the discrete, sparse, and strongly correlated nature of the features in ML-1M. Notably, \name outperforms other baselines across all datasets on both augmentation and MLE utility measures, demonstrating its suitability for tabular data synthesis on recommendations and even more various downstream tasks such as binary classification and multi-class classification.
\end{itemize}

\begin{figure}[t]
	\centering
        \subfigure[AUC.]{
		\label{fig:ab1}
		\includegraphics[width=0.45\linewidth]{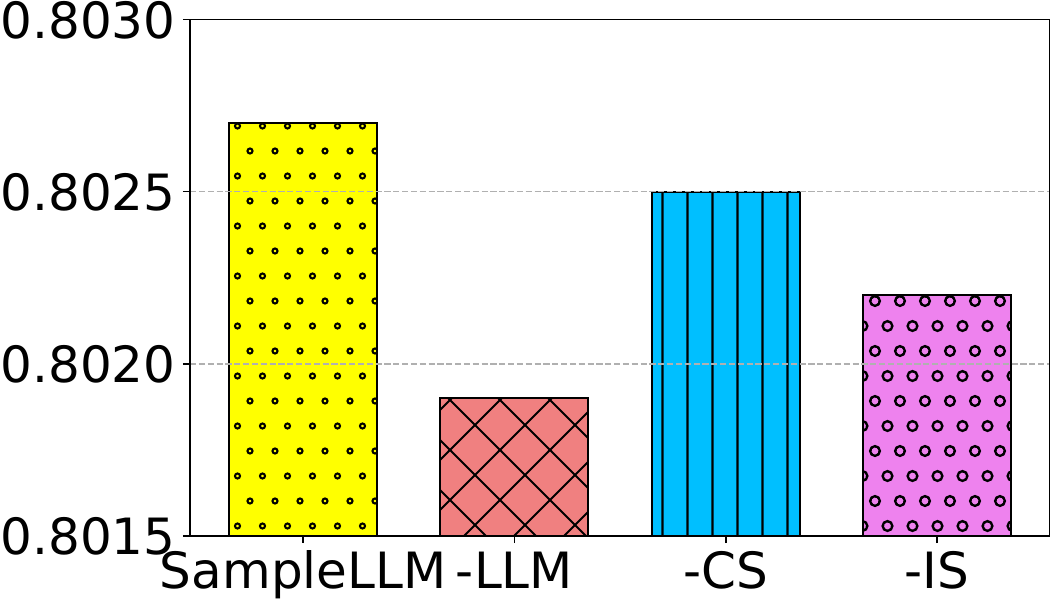}}
        \subfigure[Logloss.]{
		\label{fig:ab2}
		\includegraphics[width=0.45\linewidth]{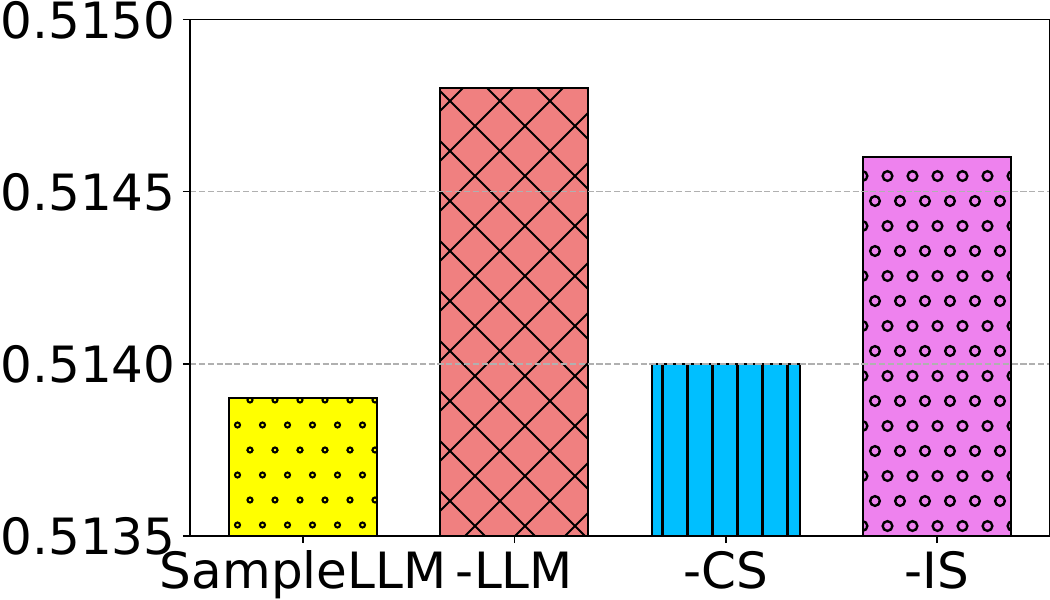}}
\vspace{-5mm}
\caption{Ablation study of augmentation utility on Douban.}
\vspace{-5mm}
\label{fig:ablation}
\end{figure}

\subsection{Ablation Study (RQ3)}
This section verifies the effectiveness of each module in \name through ablation experiments to answer RQ3. Specifically, we compare the augmentation utility of \name on the Douban dataset with the following substitutions:

\begin{itemize}[leftmargin=*]
    \item \textbf{w/o LLM (-LLM)}: with TabDDPM, a non-LLM baseline as the alternative of the LLM in the first stage of \name;
    \item \textbf{w/o cluster sampling (-CS)}: without cluster sampling in the first stage of \name. The exemplars are selected randomly.
    \item \textbf{w/o importance sampling (-IS)}: without the feature attribution-based importance sampling in the second stage of \name.
\end{itemize}

We also calculate their SDV similarity~\cite{patki2016synthetic}, a score measuring distribution similarity considering column shapes and pair trends. The result is shown in Table~\ref{tab:sdv1}.

Based on Figure~\ref{fig:ablation} and Table~\ref{tab:sdv1}, we could conclude that:

% Table generated by Excel2LaTeX from sheet 'sdv'
\begin{table}[t]
  \centering
  \caption{SDV Similarity Score with the original dataset.}
    \vspace{-4mm}
  \scalebox{0.9}{
    \begin{tabular}{ccccc}
    \toprule
    Models & SampleLLM & -LLM  & -CS & -IS \\
    \midrule
    SDV Similarity & 94.83\% & 92.32\% & 94.13\% & 93.26\% \\
    \bottomrule
    \end{tabular}%
    }
    \vspace{-4mm}
  \label{tab:sdv1}%
\end{table}%

\begin{itemize}[leftmargin=*]
    \item Replacing the LLM backbone in the first stage of \name with the TabDDPM results in a significant performance decline for \name. This decline may be attributed to LLM's superior ability to model feature relations and joint distributions at the semantic level, which allows it to better approximate the sample distribution of the original data compared to baseline methods. This characteristic is crucial for ensuring the effectiveness of feature attribution-based importance sampling in the second stage of \name, as the distribution similarity between the two datasets positively impacts the effectiveness of the importance sampling method~\cite{elvira2021advances} when the sample amount is fixed.

    \item Without the cluster sampling method in selecting exemplars at the first stage, \name also suffers from performance decline, indicating that selecting diverse exemplars is essential in LLM's few-shot learning process.

    \item Without the feature attribution-based importance sampling in the second stage, \name's performance degenerates significantly, indicating that refining sample distribution and feature relation is vital in LLM-based tabular data synthesis. 

\end{itemize}

\begin{figure}[t]
	\centering
        \subfigure[SampleLLM.]{
		\label{fig:ab0}
		\includegraphics[width=0.4\linewidth]{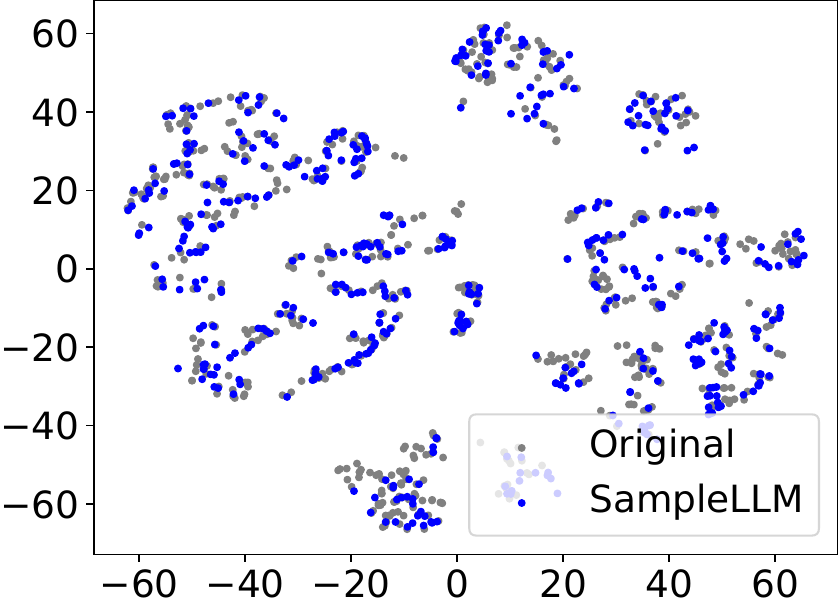}}
        \subfigure[-LLM.]{
		\label{fig:ab1v}
		\includegraphics[width=0.4\linewidth]{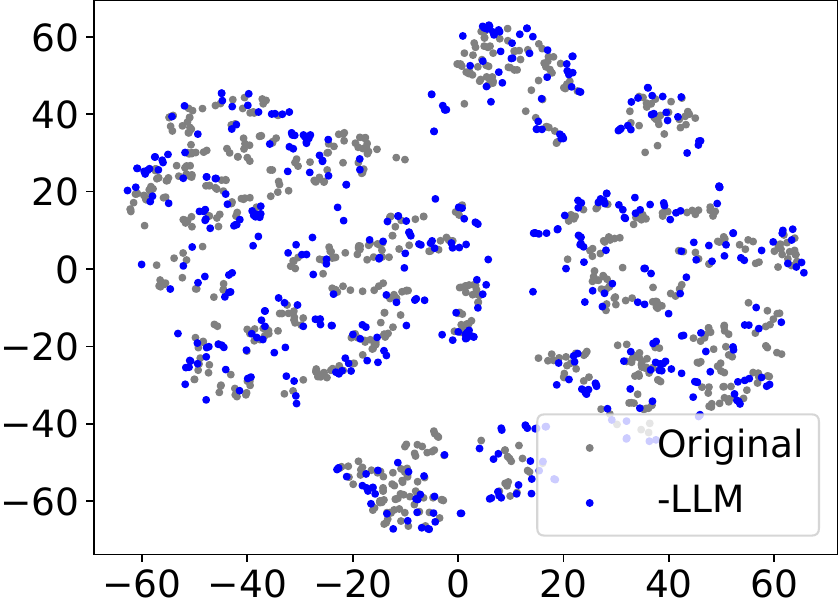}}

        \vspace{-2mm}
        \subfigure[-CS.]{
		\label{fig:ab2v}
		\includegraphics[width=0.4\linewidth]{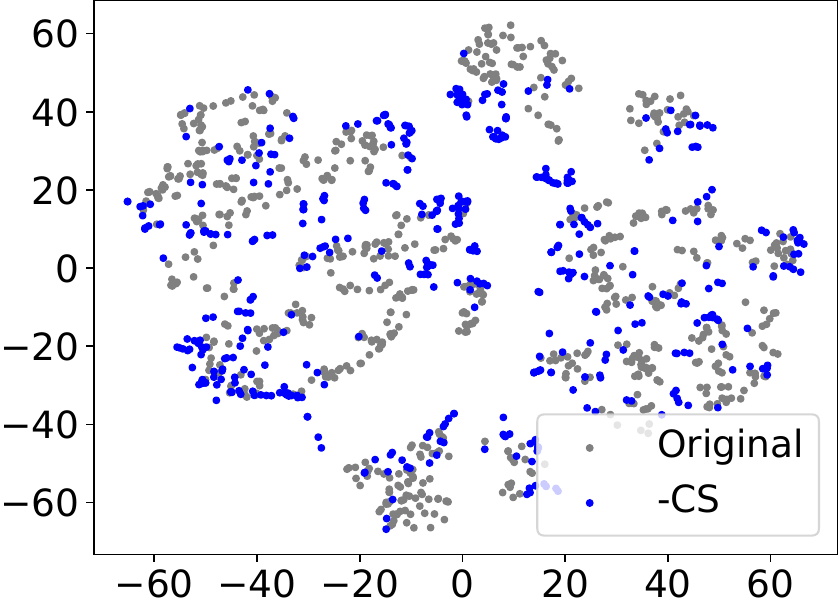}}
        \subfigure[-IS.]{
		\label{fig:ab3v}
		\includegraphics[width=0.4\linewidth]{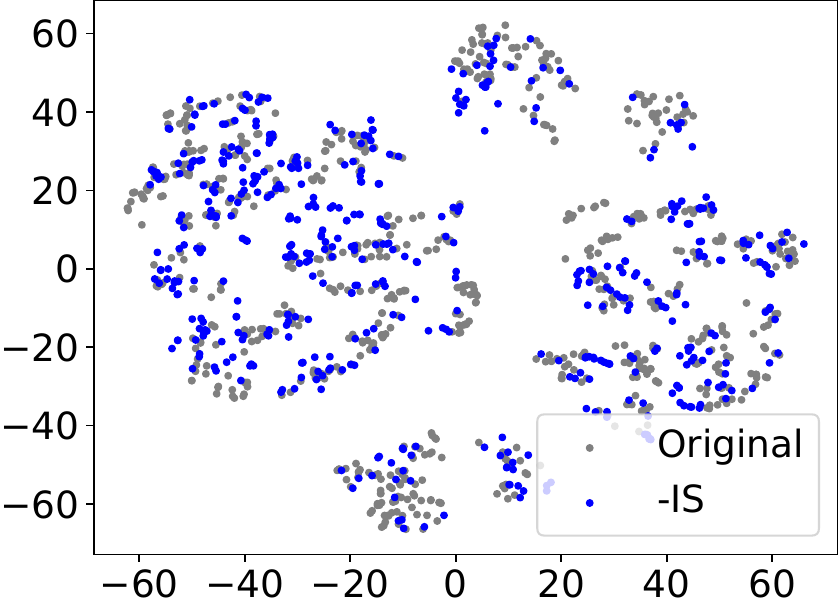}}
\vspace{-5mm}
\caption{Visualization analysis of the ablation study.}
\vspace{-5mm}
\label{fig:ablation-vis}
\end{figure}

In order to further intuitively illustrate the effectiveness of different modules, a visualization analysis is conducted through the T-Distributed Stochastic Neighbor Embedding (TSNE) method for the synthetic tabular data generated by the above alternatives, as shown in Figure~\ref{fig:ablation-vis}. From these subfigures, we could observe that:

\begin{itemize}[leftmargin=*]
    \item As depicted in Figure~\ref{fig:ab1v}, substituting the TabDDPM baseline model for the LLM in the first stage fundamentally results in a worse synthetic data distribution (especially in the middle part), despite the integration of feature attribution-based importance sampling in the subsequent stage. This observation substantiates our initial analysis.

    \item As shown in Figure~\ref{fig:ab2v}, the selection of random exemplars over those derived from cluster sampling leads to an overly concentrated synthetic data distribution in comparison with Figure~\ref{fig:ab0} when generated through the LLM.

    \item Although Figure~\ref{fig:ab3v} shows a relatively similar shape, the point density in different regions differs greatly from the original dataset. This demonstrates that, in the absence of feature attribution-based importance sampling, achieving a comprehensive alignment with the original dataset is challenging.
\end{itemize}

\section{Online A/B Test}

The Huawei app advertising platform contends with the challenge of low recall rates for cold-start apps, primarily due to the sparse user interaction data resulting from limited exposure. This scarcity hinders existing models from achieving global optimality. To address this challenge, we implemented SampleLLM to generate synthetic interactions between users and cold-start apps. In the user profile generation phase, clustering sampling was employed to iteratively select user profile seeds, which were then utilized by the LLM to synthesize diverse user profiles, simulating app exposure across a wide range of user groups. The LLM further integrated these profiles with the cold-start app's features and descriptions to predict user click behaviors, preserving the result as a raw dataset for distribution alignment.

Subsequently, we decomposed the complete feature sets into several strong interaction feature clusters using Hessian-based methods on a previously trained recommendation model. The distributions of these feature clusters were calculated based on occurrence frequency. This allows us to determine the importance weights of each synthetic sample, which serves as augmenting data to improve training for cold-start apps.

For deployment, we downsampled one day's user data and clustered it into 1,000 groups using Spark. From each cluster, samples were selected as seeds according to cluster size for subsequent data synthesis. The selected user profiles, combined with app descriptions, served as seeds for Qwen2.5 to simulate user click behaviors on the apps. The synthesized samples were then assigned importance weights based on statistics extracted from the current day's recommendation model, ensuring alignment with real user behavior for non-cold-start apps. These synthetic samples are utilized as augmentation for the following week without further updates.

We integrated the model trained with synthetic data as a new pipeline in the user-item recalling module. During the experimental phase, real cold-start samples were updated daily, while synthetic data received incremental updates weekly. \textbf{The experiment lasted a total of one week and exhibited a 2\% improvement in RPM (Revenue Per Mille), a 1.86\% improvement in ECPM (Effective Cost Per Mille)} in our cold-start scenario, thereby validating the method's effectiveness.
\section{Related Work}

Tabular data synthesis has evolved to address diverse needs across domains, initially through probabilistic models like the Synthetic Data Vault (SDV)~\cite{patki2016synthetic} that focused on privacy and statistical integrity. The introduction of deep learning approaches, including GANs~\cite{ctgan,yoon2020anonymization,jordon2018pate} and VAEs~\cite{ctgan}, has furthered this field, offering solutions to issues such as data scarcity and privacy. Models like CTGAN and TVAE~\cite{ctgan} effectively handle specific challenges, including mixed data types, significantly enhancing the quality of synthetic datasets. Specialized techniques, such as TabDDPM~\cite{kotelnikov2023tabddpm}, leverage models like denoising diffusion probabilistic models for better generative performance across benchmarks~\cite{benchmark,qian2024synthcity}. However, these methods often require large datasets, limiting their effectiveness in data-scarce contexts, especially in recommendation scenarios~\cite{zhao2020jointly,liu2023multi,lin2022adafs,li2022gromov}. Moreover, they fail to capture the semantic associations between features~\cite{zheng2022ddr,xu2024multi}, which are increasingly important in recommendation modeling.

The rise of transformer~\cite{zhao2023user} architectures~\cite{vaswani2017attention} and LLMs~\cite{zhao2023survey,minaee2024large,chang2024survey,liu2024large} has introduced new possibilities for tabular data synthesis, thanks to LLMs' few-shot learning~\cite{cahyawijaya2024llms,liu2024large1,zhang2024notellm} and semantic understanding capabilities~\cite{li2024personal}. Methods like GReaT~\cite{borisov2022language} and REaLTabFormer~\cite{solatorio2023realtabformer} have explored converting tabular data into natural language to leverage LLM strengths. Despite promising outcomes, these approaches often fail to address alignment issues between LLMs and target datasets, leading to distribution divergence. Instead, our proposed method, \name, addresses these shortcomings by combining few-shot learning with alignment techniques to better match distributions and feature relationships of generated and original datasets, thereby enhancing data utility and downstream task performance in recommendations.
\section{Conclusion}
In this paper, a two-stage LLM-based framework \name is proposed to integrate LLM with sampling methods for optimizing tabular data synthesis in recommendations. Specifically, a manually designed instruction, together with a group of exemplars generated through a cluster sampling method serves as the input prompt for few-shot learning via LLM. Then a novel feature attribution-based importance sampling method is proposed to serve as a second stage for further feature relation modeling and distribution alignment. By doing so, \name is able to generate synthetic tabular data with semantic understanding, higher utility, and distribution similarity. Experiments on three public recommendation datasets, two general datasets, and an online application demonstrate the effectiveness of the proposed \name.

\begin{acks}
This research was partially supported by Research Impact Fund (No.R1015-23), Collaborative Research Fund (No.C1043-24GF), APRC - CityU New Research Initiatives (No.9610565, Start-up Grant for New Faculty of CityU), Hong Kong ITC Innovation and Technology Fund Midstream Research Programme for Universities Project (No.ITS/034/22MS), and Huawei (Huawei Innovation Research Program).
\end{acks}

%% The acknowledgments section is defined using the "acks" environment
%% (and NOT an unnumbered section). This ensures the proper
%% identification of the section in the article metadata, and the
%% consistent spelling of the heading.
% \begin{acks}
% xxx.
% \end{acks}

%%
%% The next two lines define the bibliography style to be used, and
%% the bibliography file.
\clearpage
\normalem
\bibliographystyle{ACM-Reference-Format}
\bibliography{bibfilenew}

%%
%% If your work has an appendix, this is the place to put it.
\clearpage
\appendix

\section{Pseudo-code of \name}\label{sec:app-pseudo}
To elucidate the overall process of \name, we present the pseudo-code in Algorithm~\ref{alg:overall}. Specifically, \name operates in two distinct stages. The first stage (i.e., lines 1-7) employs a cluster sampling technique for exemplar selection in LLM's few-shot learning, aimed at generating an initial synthetic tabular dataset, denoted as $\mathcal{D}_{s'}$. In the second stage (i.e., lines 8-12), a novel feature attribution-based importance sampling method is proposed to enhance the modeling of feature relationships and achieve distribution alignment between the synthetic dataset $\mathcal{D}_{s'}$ and the original tabular dataset $\mathcal{D}_o$. This process results in the final synthetic tabular dataset, $\mathcal{D}_{s}$.

\begin{algorithm}[t]
	\caption{\label{alg:overall} The whole process of \name}
	\raggedright
	{\bf Input}:  The original tabular dataset $\mathcal{D}_o$; An LLM model $LLM$; Hyper-parameters $a$, $b$, $Q$, $\gamma$\\
	{\bf Output}: The final synthetic tabular dataset $\mathcal{D}_s$\\

	\begin{algorithmic} [1]
        \STATE $\mathcal{D}_{s'} = []$
        \FOR{Iteration in 1,..., $Q$}
    	    \STATE Obtain instruction $I$ for $\mathcal{D}_o$ through Section~\ref{sec:instruction}
            \STATE Obtain exemplar set $E$ with $a$ exemplars via Equation~\eqref{equ:exemplar}
            \STATE Obtain output synthetic data $\mathcal{D}_{tmp}$ with $b$ samples via $LLM(I+E)$
            \STATE $\mathcal{D}_{s'} = \mathcal{D}_{s'} + \mathcal{D}_{tmp}$
        \ENDFOR
        \STATE Obtain $\boldsymbol{\Gamma}^{\mathcal{D}_o}$ via Equation~\eqref{equ:gamma_1}, ~\eqref{equ:gamma_2}
        \STATE Obtain feature groups $Gp$ via Equation~\eqref{equ:group}
        \STATE Obtain discriminant probability $\mathcal{M}_o$ and $\mathcal{M}_{s'}$ via Equation~\eqref{equ:probability}
        \STATE Obtain sample weight $w$ for samples in $\mathcal{D}_{s'}$ via Equation~\eqref{equ:weights}
        \STATE Obtain $\mathcal{D}_{s}$ by processing importance sampling on $\mathcal{D}_{s'}$ with sample weight $w$.

	\end{algorithmic}

\end{algorithm}

\section{Overall Interaction Map and Feature Groups}\label{sec:app-map}

\begin{figure}[t]
	\centering
        \subfigure[ML-1M.]{
		\label{fig:inter-ml-1m}
		\includegraphics[width=0.34\linewidth]{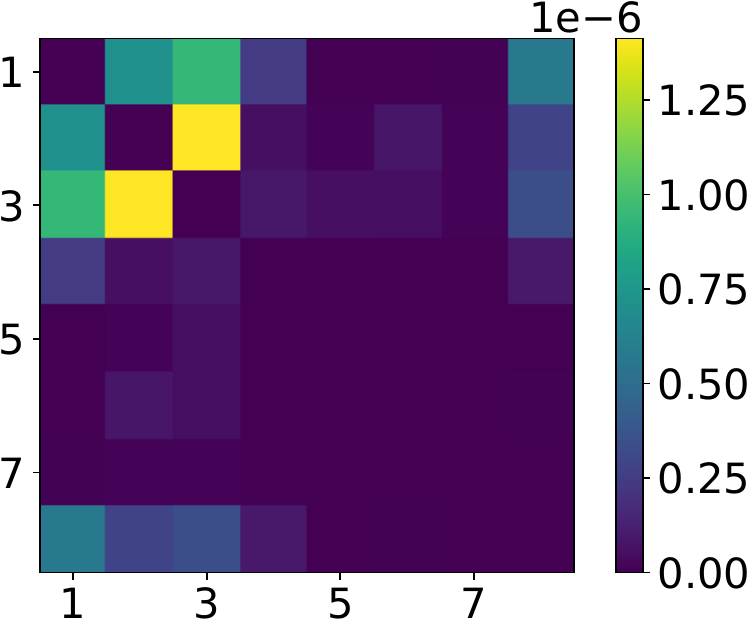}}
        \subfigure[Amazon.]{
		\label{fig:inter-amazon}
		\includegraphics[width=0.3\linewidth]{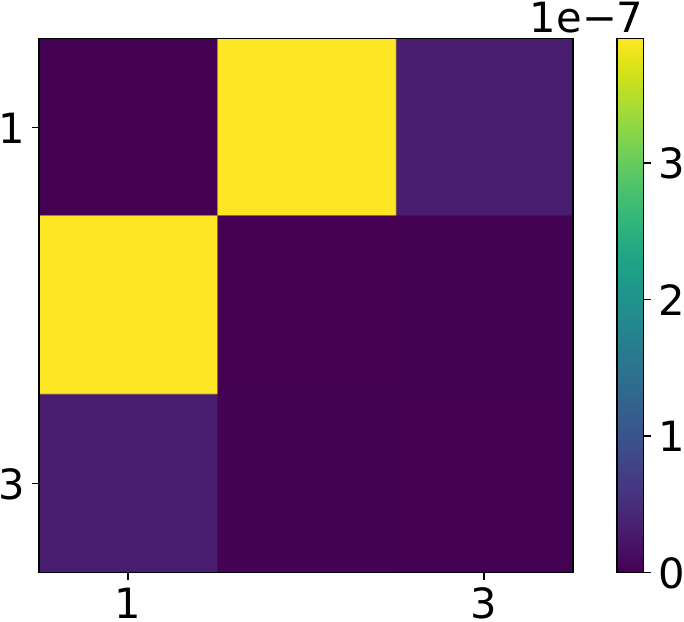}}

        \vspace{-3mm}
        \subfigure[Douban.]{
		\label{fig:inter-douban}
		\includegraphics[width=0.31\linewidth]{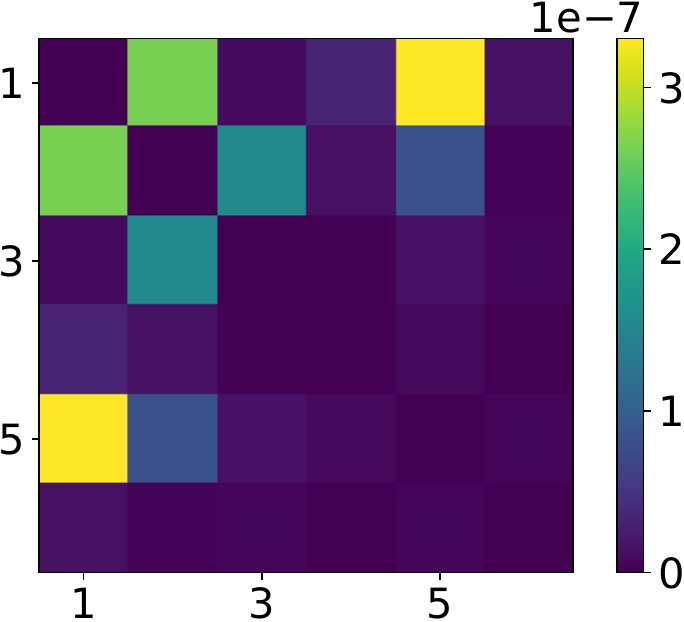}}
        \subfigure[HELOC.]{
		\label{fig:inter-heloc}
		\includegraphics[width=0.32\linewidth]{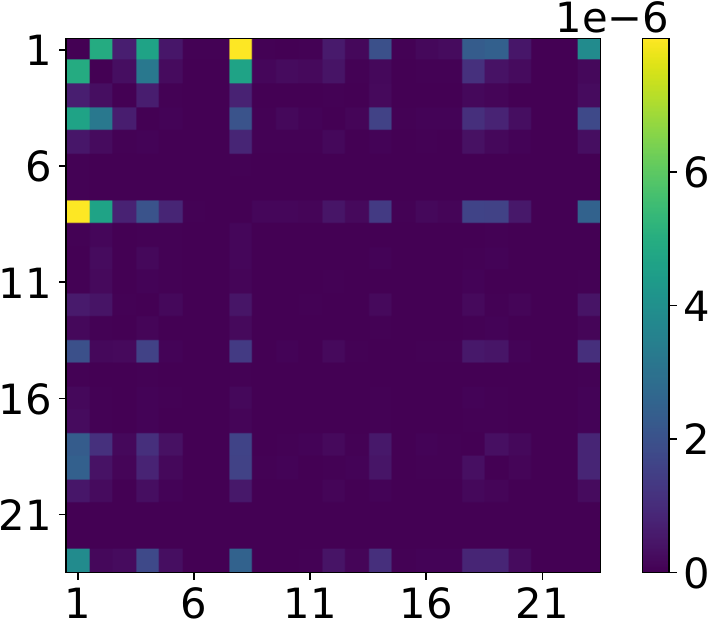}}

        \vspace{-3mm}
        \subfigure[Covertype.]{
		\label{fig:inter-covertype}
		\includegraphics[width=0.4\linewidth]{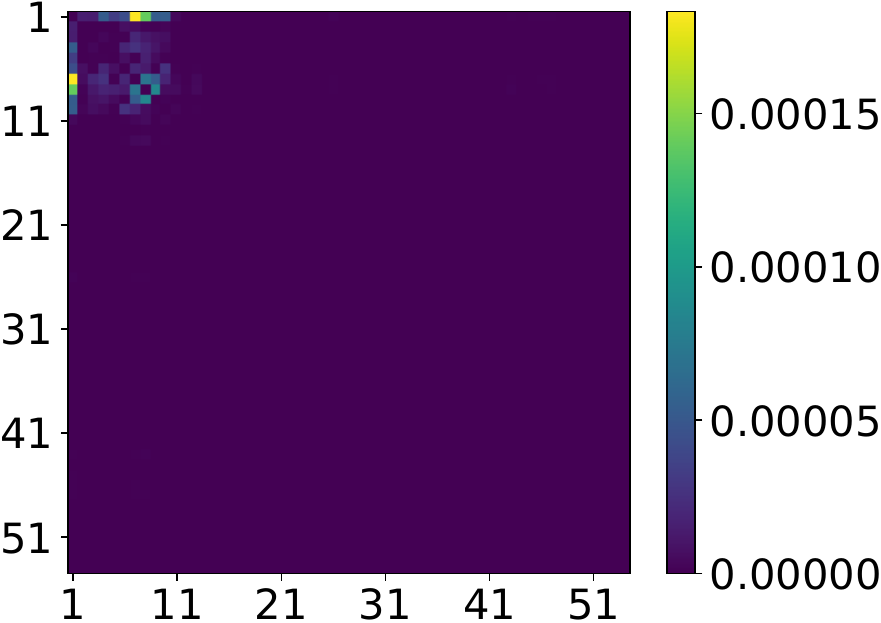}}
\vspace{-4mm}
\caption{Interaction maps for different datasets. The numbers on the axes represent the indices of the feature fields.}
\vspace{-3mm}
\label{fig:intermap}
\end{figure}

In this section, we present the overall interaction maps and feature groups for all five experimental datasets, as illustrated in Figure~\ref{fig:intermap}, and Table~\ref{tab:groups}. It is important to note that the interaction maps and feature groups may exhibit slight variations from the provided figures and tables due to fluctuations in the performance of the predictive model trained on each dataset.

% Table generated by Excel2LaTeX from sheet 'overall-1'
\begin{table}[t]
  \centering
  \vspace{-3mm}
  \caption{Feature groups. Note that only groups with more than one feature field are shown in the table for simplicity.}
  \vspace{-3mm}
  \scalebox{0.9}{
    \begin{tabular}{cc}
    \toprule
    Dataset & Feature Group \\
    \midrule
    ML-1M & [1, 2, 3] \\
    Amazon & [1, 2] \\
    Douban & [1, 2, 5] \\
    HELOC & [1, 8] \\
    Covertype & [1, 7, 8] \\
    \bottomrule
    \end{tabular}%
    }
    \vspace{-3mm}
  \label{tab:groups}%
\end{table}%

\section{Case Study}\label{sec:case}

\begin{figure}[t]
	\centering
        \subfigure[SampleLLM.]{
		\label{fig:case0}
		\includegraphics[width=0.35\linewidth]{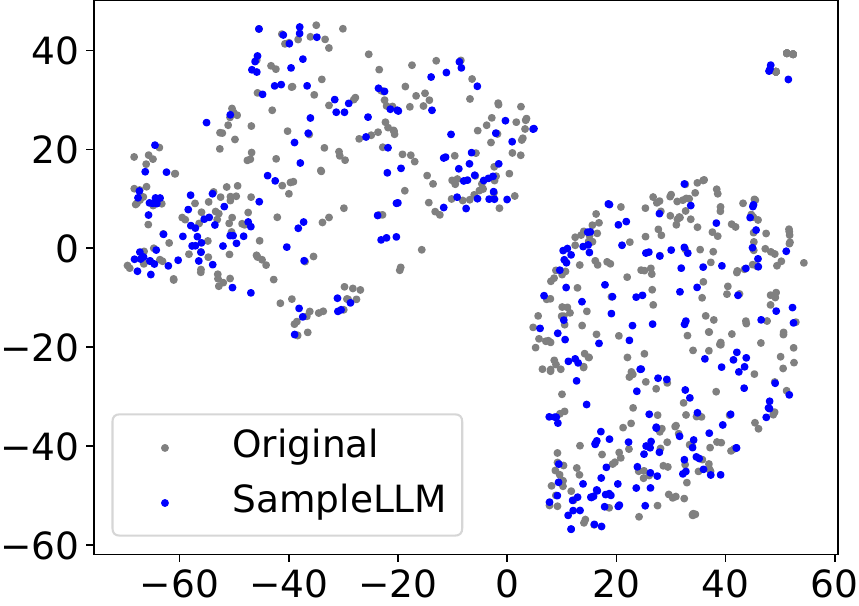}}
        \subfigure[TVAE.]{
		\label{fig:case1}
		\includegraphics[width=0.35\linewidth]{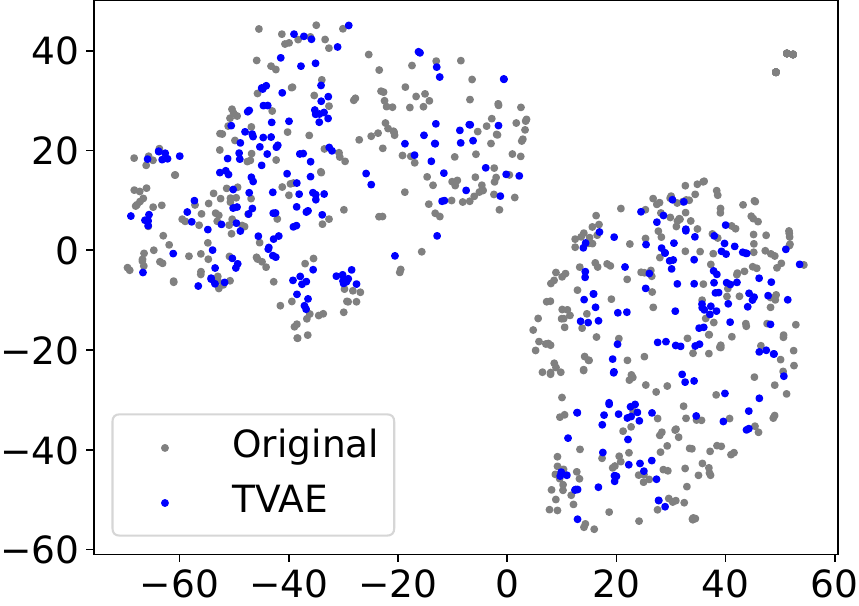}}
        
        \vspace{-2mm}
        \subfigure[TabDDPM.]{
		\label{fig:case2}
		\includegraphics[width=0.35\linewidth]{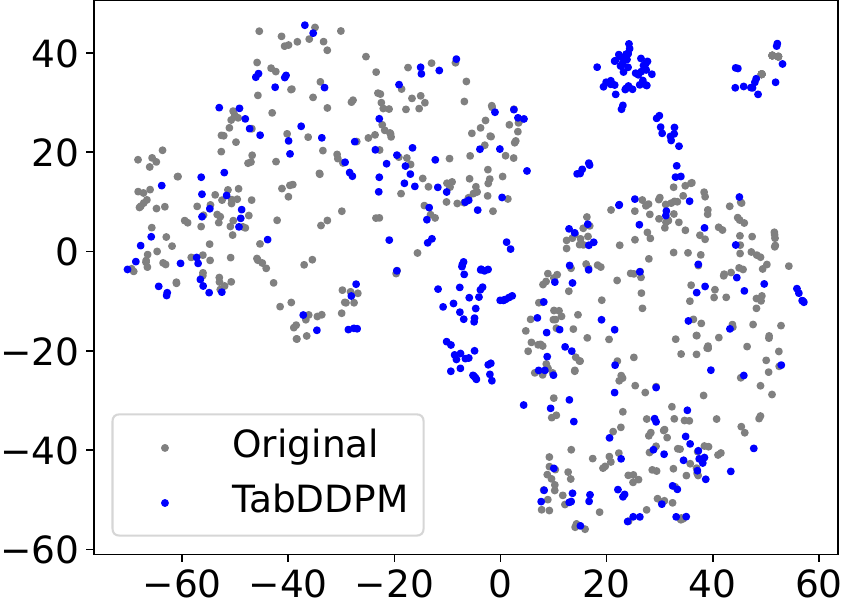}}
        \subfigure[REaLTabFormer.]{
		\label{fig:case3}
		\includegraphics[width=0.35\linewidth]{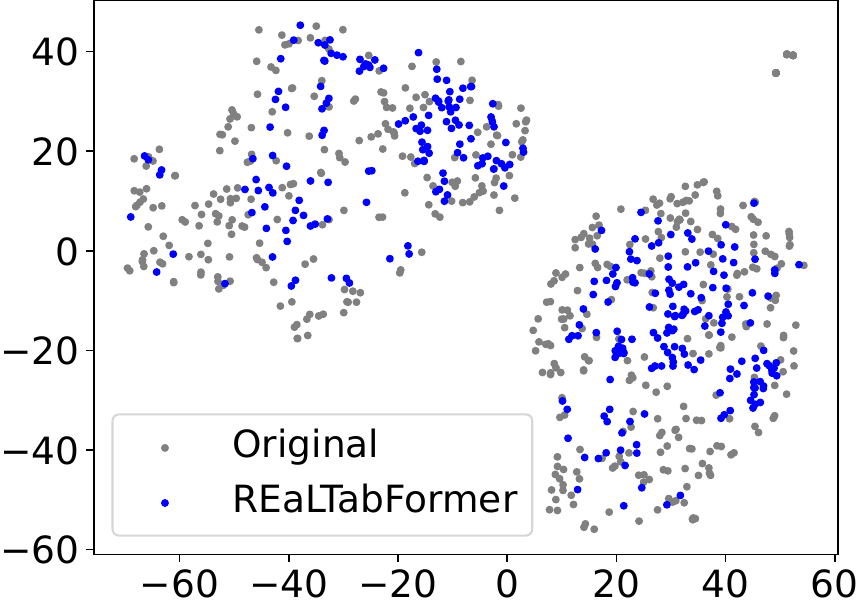}}
        \vspace{-2mm}
\caption{Case study on HELOC dataset.}
\vspace{-2mm}
\label{fig:case}
\end{figure}

This section presents a case study using TSNE visualizations and SDV similarity~\cite{patki2016synthetic}, a score measuring distribution similarity considering column shapes and pair trends, to analyze the synthetic tabular data generated by high-performance baselines such as TVAE, TabDDPM, and REaLTabFormer, thereby highlighting the advantages of \name. As shown in Figure~\ref{fig:case} and Table~\ref{tab:sdv}, TVAE and TabDDPM struggle to effectively replicate the original data distribution. Even with the incorporation of LLM, REaLTabFormer's distribution remains concentrated in several central regions, as noted in Figure~\ref{fig:introduction}. In contrast, \name demonstrates significant improvements in aligning with the original sample distribution by leveraging the few-shot learning capabilities of LLM and employing a novel two-stage alignment strategy.

% Table generated by Excel2LaTeX from sheet 'sdv'
\begin{table}[t]
  \centering
  \caption{SDV Similarity Score with the original dataset.}
  \vspace{-2mm}
  \scalebox{0.85}{
    \begin{tabular}{ccccc}
    \toprule
    Models & SampleLLM & TVAE  & TabDDPM & REaLTabFormer \\
    \midrule
    SDV Similarity & 90.10\% & 85.96\% & 84.84\% & 87.23\% \\
    \bottomrule
    \end{tabular}%
    }
    \vspace{-2mm}
  \label{tab:sdv}%
\end{table}%

\clearpage

\end{document}